\documentclass{ws-procs9x6}

\def\be{\begin{equation}}
\def\ee{\end{equation}}
\newcommand{\bea}{\begin{eqnarray}}
\newcommand{\eea}{\end{eqnarray}}

\newcommand{\p}{{\bf p}}

\begin{document}

\title{The nature of the soft excitation at the critical end point of QCD
\footnote{\uppercase{T}his work is supported by 
the \uppercase{H}ungarian \uppercase{R}esearch \uppercase{F}und
(\uppercase{OTKA}) under contract numbers \uppercase{F}043465,
\uppercase{T}034980, \uppercase{T}037689, and \uppercase{T}046129.}}

\author{A. Jakov\'ac}

\address{
HAS \footnote{\uppercase{H}ungarian \uppercase{A}cademy of 
\uppercase{S}ciences} and Budapest Univ. of
Technology and Economics, Research Group ``Theory of Condensed Matter'',
H-1521 Budapest, Hungary
}

\author{A. Patk{\'o}s}

\address{Department of Atomic Physics, E{\"o}tv{\"o}s University,
H-1117 Budapest
}  

\author{Zs. Sz{\'e}p}

\address{
Research Group for Statistical Physics, HAS, H-1117 Budapest\\
E-mail: szepzs@antonius.elte.hu
}

\author{P. Sz{\'e}pfalusy}

\address{
Dept. of Physics of Complex Systems,
E{\"o}tv{\"o}s University, H-1117 Budapest\\
Research Inst. for Solid State Physics and Optics,
HAS, H-1525 Budapest
}

\maketitle

\abstracts{
Using a large flavor number expansion and a gap equation for the pion mass
the chiral quark-meson model is solved at the lowest order in the fermion
contributions. In the chiral limit the tricritical point (TCP) is
determined analytically. The softening of the sigma particle is verified 
at this point and is further investigated for a physical pion mass in the 
neighbourhood of the critical endpoint (CEP).
}

\section{Introduction}

Recently the location of the CEP in the $\mu-T$ phase-diagram of QCD was
obtained for $N_f=2+1$ dynamical staggered quarks with physical
masses~\cite{katz04}. One might investigate this problem using an effective
low energy model also, with some nonperturbative technique. Its low energy
nature might prevent us from giving the accurate location of the endpoint
and the shape of the phase boundary, but one can still obtain insights into
the physics near the CEP. In this contribution we focus on the nature of the
soft mode.

\section{The method for solving the model}

In the presence of the vacuum expectation value $\Phi$ the
requirement of a finite constituent quark mass
$m_q=g\Phi$ as $N\rightarrow \infty$ defines the Lagrangian  
\bea
L[\sigma,\pi^a,\psi]=
-\left[\frac{\lambda}{24}\Phi^4+\frac{1}{2}m^2\Phi^2\right]N-
\left[\frac{\lambda}{6}\Phi^3+m^2\Phi-h\right]\sigma\sqrt{N}
\nonumber\\
+\frac{1}{2}\bigl[(\partial\sigma)^2 + (\partial\vec\pi)^2 \bigr]
-\frac{1}{2}m^2_{\sigma 0}\sigma^2
-\frac{1}{2}m^2_{\pi 0}\vec\pi^2
-\frac{\lambda}{6\sqrt{N}}\Phi\sigma \rho^2-
\frac{\lambda}{24N}\rho^4 \nonumber\\
+\bar \psi\left[i\partial^\mu\gamma_\mu-m_q-\frac{g}{\sqrt{N}}
\left(\sigma
+i\sqrt{2N_f}\gamma_5T^a\pi^a\right)\right]\psi
,
\eea
where $N=N_f^2,$ $m^2_{\sigma 0}=m^2+\frac{\lambda}{2}\Phi^2$,
$m^2_{\pi 0}=m^2+\frac{\lambda}{6}\Phi^2$,
$\rho^2=\sigma^2+\vec\pi^2$ and $h$ is the external field.
The fermions, for which the chemical potential is introduced by the shift
$\partial_t\rightarrow \partial_t-i\mu$, start contributing
in the large N expansion at level $\mathcal{O}(1/\sqrt{N})$, 
which is between LO and NLO of the mesonic sector. 
To calculate their corrections to the LO, which has O(N) symmetry,
we disregarded the second independent quartic coupling of the
model, present for $N_f> 2$. 

Due to the proliferation of graphs a resummation is needed. The best thing
to do would be to use a selfconsistent pion and fermion propagator. This
makes renormalisation difficult, and a consistent method for it was proposed
only recently \cite{Reinosa}. Our method consists in taking into account the
$N_c$ coloured quarks to ${\mathcal O}(1/\sqrt{N})$ perturbatively at one
loop order and using for the propagator of the pions a simplified form,
parameterised by a mass M determined from the gap-equation:
\be
M^2=-iG_\pi^{-1}(p=0)=
m^2+\frac{\lambda}{6}\Phi^2+\frac{\lambda}{6}T_B(M)-4\frac{g^2N_c}{\sqrt{N}}
T_F(m_q).
\label{Eq:gap}
\ee
The pion and fermion tadpoles ($T_{B,F}$) determine  the equation of state (EoS)
\be
V'_\textrm{eff}=\Phi H(\Phi^2)-h=
\Phi\left[m^2+\frac{\lambda}{6}\left(\Phi^2+T_B(M)\right)
-\frac{4g^2N_c}{\sqrt{N}}T_F(m_q)\right]-h=0.
\label{Eq:EoS}
\ee  
The renormalisation method was presented in \cite{Toni_renorm}. 
The relation $M^2=H(\Phi^2)$ shows that in the symmetry broken 
regime defined by $H(\Phi^2)=h/\Phi$ and in the chiral limit ($h=0$) the 
physical pion mass $M$ is always zero, i.e. Goldstone's theorem is fulfilled 
in the minimum of the effective potential.
In what follows the number of flavors is set to two: $N_f=2$.

\section{Chiral case ($h=0$): analytical determination of TCP}
\label{sec:analysys}

\begin{figure}[ht]
\centerline{\epsfxsize=5.95cm\epsfbox{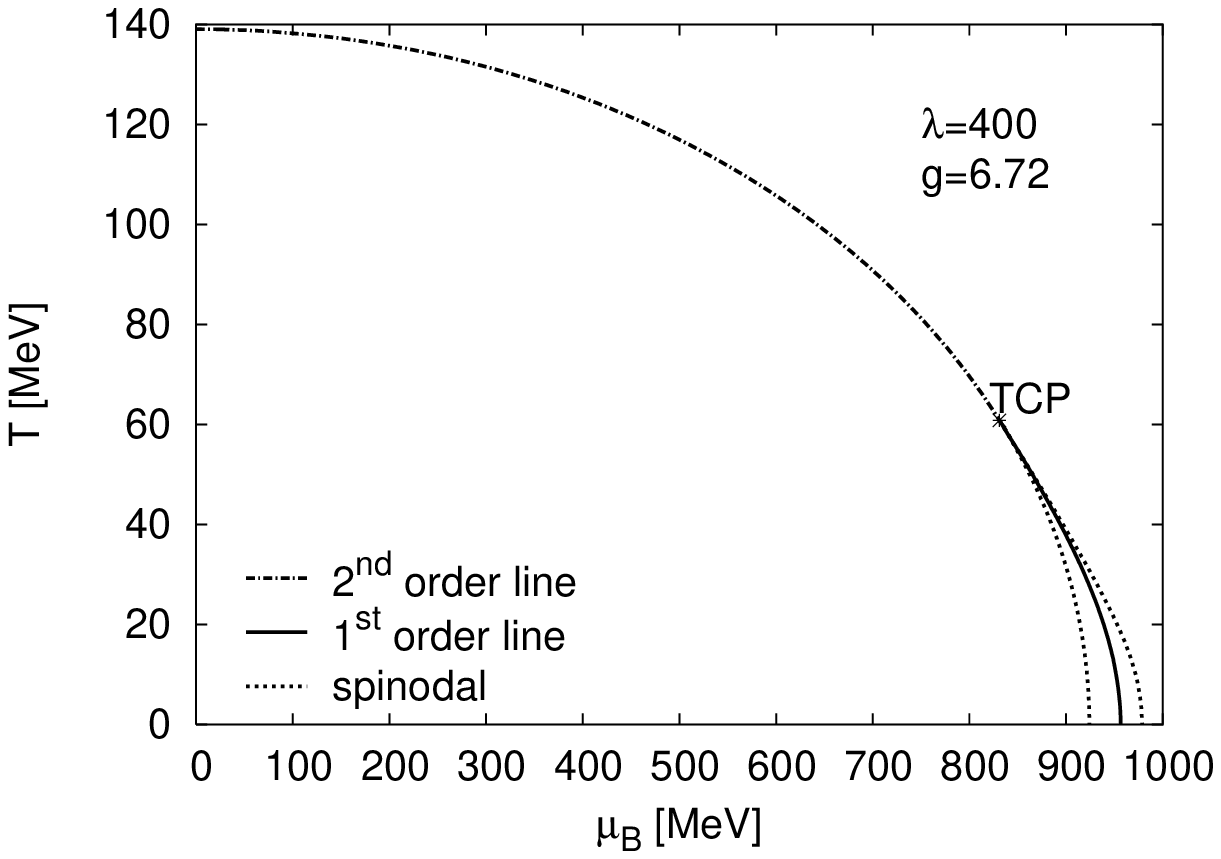}
\hspace*{-0.3cm}
\epsfxsize=5.95cm\epsfbox{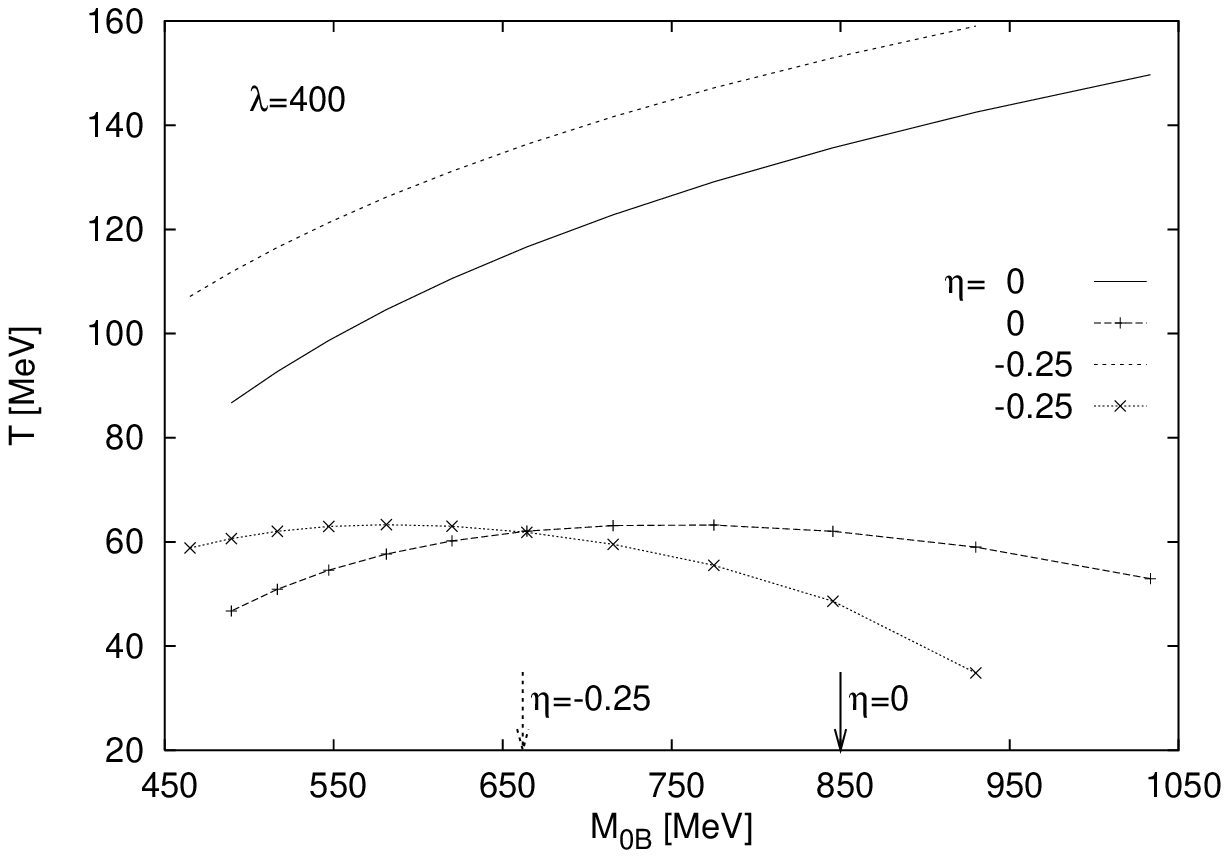}}
\caption{L.h.s.: The $T-\mu$ phase diagram for $M_{0B}=886$ MeV and
$\eta=0$. R.h.s.  the dependence of $T_c(\mu =0)$ (upper curves) and of
$T_{TCP}$ (lower curves) on $M_{0B}$.
\label{Fig}}
\end{figure}

Since in this case $M=0$, one obtains \cite{QMM03} the phase boundary by 
expanding the fermion tadpole in powers of $\Phi$. The 2$^{\textrm{nd}}$ 
order line is given by
\be
m_T^2:=m^2+\frac{g^2\mu^2}{4\pi^2}N_c+
\left(\frac{\lambda}{72}+\frac{g^2N_c}{12}\right)T^2=0,
\label{Eq:quadrat}
\ee
while the location of TCP comes from combining (\ref{Eq:quadrat}) with the 
condition
\be
\frac{\bar\lambda}{6}:=
\frac{\lambda}{6}+\frac{g^4N_c}{4\pi^2}\left[\frac{\partial}{\partial n}
\Big(\textrm{Li}_n(-e^{-\frac{\mu}{T}})+\textrm{Li}_n(-e^{-\frac{\mu}{T}})
\Big)\Big|_{n=0}- \ln\frac{c_1T}{M_{0B}}\right]=0.
\label{Eq:quartic}
\ee
$\ln c_1/2=1-\gamma_E+\eta$, and $\eta=\ln M_{0B}/M_{0F}$ 
gives a relation between $M_{0B}$ and $M_{0F}$ the renormalisation scales 
of the bosonic and fermionic tadpoles. The parameters 
are fixed at $T=\mu=0$. 
The choice $m_{q0}=m_N/3\approx 312.67$ MeV and $\Phi_0=f_\pi/2$ gives 
$g=6.72$. $\lambda=400$ is fixed by requiring the agreement between the 
location of the complex $\sigma$-pole on 
the 2$^{\textrm{nd}}$ Riemann sheet and the experimentally 
favoured mass and width of the $\sigma$ particle. Solving 
(\ref{Eq:quadrat}) and (\ref{Eq:quartic}) the 2$^{\textrm{nd}}$ order 
phase transition, the first
spinodal line and the location of TCP are found 
(l.h.s. of Fig.~\ref{Fig}).
On the r.h.s.  we see the variation of TCP and of the critical temperature with
$\eta$ and $M_{0B}$. The pole structure of the $\sigma$
propagator restricts the allowed range of $M_{0B}$ to the right from the
arrow, seen on the horizontal axis, where no low scale
imaginary poles appear besides the well known large scale tachyon, whose
scale increases with decreasing values of 
$\lambda$. 
Although in this region one can easily obtain $T_c(\mu=0)\in(150,170)$ MeV, 
the TCP stays robustly below $70$~MeV. Previous 
effective model studies \cite{LEeff} give similar low
values compared with the CEP temperature  obtained in~\cite{katz04}, 
which could mean that the phase transition is 
driven by higher excited hadronic states rather than by
the light d.o.f.
This is supported by a resonance gas model calculation \cite{Redlich}
used to reproduce lattice results 
on quark number susceptibilities and pressure.

An approximate effective potential is obtained by integrating the EoS 
away from equilibrium and subsequently Taylor expanding it around 
$\Phi=0$. This value becomes a minimum as $T\to T_c$ and so $M\to 0$.
With this, one can show the vanishing of the effective coupling in the 
large N limit \cite{Wetterich} at the critical point.
Here we just quote the result of \cite{Bp04}
\be
V_\mathrm{approx}(\Phi)= \frac{m_T^2}{4U^2W}\left[
\frac{m_T^2}2\Phi^2 +\frac{\bar\lambda}{12} \Phi^4+
\left(\frac{W\bar\lambda^2}{216m_T^2}+\frac{\kappa}{3}\right)\Phi^6
\right],
\label{eq:Veff_Tc}
\ee
where 
$\kappa=\frac{g^6N_c}{(4\pi T)^2\sqrt{N}}\frac{\partial}{\partial n}
\left[\mathrm{Li}_n(-e^{\frac{\mu}{T}})+ \mathrm{Li}_n(-e^{-\frac{\mu}{T}})
\right]\big|_{n=-2}$,
$W=1-\frac{\lambda^2}{48\pi^2}\times\ln\frac{c_2T}{M_0}$ and
$\ln \frac{c_2}{4\pi} = \frac{1}{2}-\gamma$. 
The shape of $V_\mathrm{approx}(\Phi)$
depends on the 
values of the 
parameters. In the temperature range of a $2^{\textrm{nd}}$ order 
phase transition $W<0$. For increasing values of $\mu/T$, $\kappa$ 
changes sign at 1.91.
This sign change restricts the location of the
TCP to the region $\mu>2T$, where $\kappa>0$.
In the broken symmetry phase $m_T^2<0$ and it vanishes at $T_c$
signalling that the coefficient of the quartic term in $\Phi$ also vanishes. 
Around $T_c$ from (\ref{Eq:quadrat}) one gets $m_T^2\sim T-T_c$, which 
through $m_\mathrm{eff}=\big|\frac{m_{T}^2}{2U W}\big|$
gives the correct LO critical exponent $\nu=1$ for the O(N)
model at large $N$. 
The usual Landau type analysis applies to 
the square bracket in (\ref{eq:Veff_Tc}). 
The scaling exponent of the order parameter on the $2^{\textrm{nd}}$ order 
line is given by the first two terms. The minimum condition gives
$\Phi^2=-\frac{3m_T^2}{\bar\lambda}\,\Rightarrow
\Phi\sim(T-T_c)^{\beta}$, $\beta=\frac{1}{2}$. At TCP one can 
set $\bar\lambda=0$ in (\ref{eq:Veff_Tc}) and keep the sixth order term. The
minimum condition gives
$\Phi^4=-\frac{m_T^2}{2\kappa}\,\Rightarrow \Phi\sim(T-T_c)^{\beta}$, 
$\beta=\frac{1}{4}$. This mean field estimates for $\beta$ were checked numerically
using the exact EoS (\ref{Eq:EoS}).

\section{Case of the physical pion mass: the soft mode at CEP}

In view of the gap-equation (\ref{Eq:gap}) and EoS (\ref{Eq:EoS}) the form of 
the $\sigma$ propagator can be inferred from the consistency condition
$-G_\sigma^{-1}(p=0)=\frac{d^2 V_{eff}(\phi)}{d\Phi^2}$:
\be
G_\sigma^{-1}(p)=p^2-\frac{h}{\phi}-
2\Phi^2\frac{\frac{\lambda}{6}-\frac{4g^4N_c}{\sqrt{N}} I_F(p,m_q)}
{1-\frac{\lambda}{6} I_B(p,M)},
\ee
where $I_{B,F}$ are the pion and fermion one-loop ``fish'' integrals. 
Requiring $M=140$ MeV at $T=\mu=0$ one has from (\ref{Eq:gap}) 
$h/f_\pi^3=1.13$. The location of CEP is determined numerically by 
simultaneously solving (\ref{Eq:gap}) and (\ref{Eq:EoS}): 
$(\mu_B,T)_\textrm{CEP}=(987,12.34)$ MeV. $T_{CEP}$ is unrealistically low, 
nevertheless it is of interest to study which mode becomes soft at this point.
Two minima meet at CEP which flattens the effective potential. This is shown 
also by the diverging peak of the susceptibility $d\Phi/d h$. This static 
information has to show up consistently when investigating 
$G_\sigma(p)$
in the static limit {\it i.e.} first taking $p_0\to 0$, then 
$|\p|\to 0$. The question is in which $(p_0,|\p|)$ 
region one has to look for  the responsible excitations. 
The behaviour of the spectral function 
$\rho(p_0,|{\bf p}|)=-\frac{1}{\pi}
\underset{\varepsilon\to 0}{\lim}\,\textrm{Im}\,
G_\sigma(p_0+i\varepsilon,|{\bf p}|)$ 
for fixed $|\p|$ gives a hint in this respect:
a peak is found in the $p_0<|\p|$ region. The location of 
its maximum moves toward the origin with decreasing values of $|\p|$ and 
at CEP eventually the peak diverges at $p_0=0$ as $|\p|\to 0$. The nature of
the analytical object on the second Riemann sheet which produces this
peak will be explored further. In the NJL model  a pole on the negative 
imaginary axis, approaching the origin as $|\p|\to 0$, is responsible for 
the peak observed in  the spectral function \cite{Fujii}.

\section{Conclusions}

Contrary to the chiral case, where the zero temperature $\sigma$ pole
continuously moves to the origin of the 2$^\textrm{nd}$ Riemann sheet and
becomes soft at TCP, for a physical pion mass the $\sigma$ pole stays
massive at CEP. A space-like $(0\le p_0<|{\bf p}|)$ excitation is responsible
for the flattening of the effective potential at this point. Before a more
detailed investigation of the analytic structure at CEP we have to improve
our method of solving the model by dynamically generating the fermion mass
through a Dyson-Schwinger equation for the fermion propagator, but also one
has to extend our method to the physically more appealing 
$SU(3)_L\times SU(3)_R$ case.

%
%
%
%


\begin{thebibliography}{0}

\bibitem{katz04}Z. Fodor, S. D. Katz, {\it JHEP} {\bf 0404} (2004) 050
\bibitem{Reinosa} J.-P. Blaizot {\it et al.}, {\it Nucl. Phys.} 
{\bf A 736} (2004) 149, see also this  proceedings
\bibitem{Toni_renorm}A. Jakov\'ac, Zs. Sz{\'e}p, {\it hep-ph/0405226},
see also this  proceedings
\bibitem{QMM03} A. Jakov\'ac, A. Patk{\'o}s, Zs. Sz{\'e}p, P. Sz{\'e}pfalusy,
{\it Phys. Lett.} {\bf B582} (2004) 179
\bibitem{LEeff}J. Berges and K. Rajagopal, {\it Nucl. Phys.}
{\bf B538} (1999) 215; 
O. Scavenius {\it et al.}, {\it Phys. Rev.} {\bf C64} (2001) 045202; 
B.-J. Schaefer {\it et al.}, {\it nucl-th/0403039}; A.~Barducci {\it et al.},
{\it Phys. Rev.} {\bf D69} (2004) 096004
\bibitem{Redlich} F. Karsch, K. Redlich, A. Tawfik, {\it Eur. Phys. J.}
{\bf C29} (2003) 549
\bibitem{Wetterich}M. Reuter, N. Tetradis, C. Wetterich,
{\it Nucl. Phys.} {\bf B401} (1993) 567
\bibitem{Bp04}A. Jakov\'ac, A. Patk{\'o}s, Zs. Sz{\'e}p, P. Sz{\'e}pfalusy,
{\it hep-ph/0406122}
\bibitem{Fujii}H. Fujii, {\it Phys. Rev.} {\bf D67} (2003) 094018 
\end{thebibliography}
\end{document}